%% file: 00_Main_SichER.tex
\documentclass[review,preprint,12pt,nonatbib,english]{elsarticle}
\usepackage[margin=3cm]{geometry}

\usepackage[utf8]{inputenc}
\usepackage[T1]{fontenc}
\usepackage{babel}[english]

\usepackage{crimson}

\usepackage{amssymb}
\usepackage{amsmath}
\usepackage{siunitx}
\usepackage{url}
\usepackage{hyperref}
\hypersetup{breaklinks=true}
\usepackage[version=4]{mhchem}
\usepackage{eurosym}
\usepackage[nolist]{acronym} 			
\usepackage{booktabs}	
\usepackage{cleveref}

\usepackage{graphicx}
\usepackage{multirow,multicol,booktabs,makecell}
\usepackage{threeparttable}
\usepackage{longtable}
\usepackage{rotating}
\usepackage{subfig}
\usepackage{pdfpages}
\usepackage{listings}
\usepackage{array}
\usepackage{tabularx}
\usepackage{tabulary}
\usepackage{ragged2e}
\usepackage{xcolor}
\usepackage{soul}
\usepackage{setspace}
\usepackage{pdflscape}
\usepackage{float}
\usepackage{csquotes}

\usepackage{footnote}
\makesavenoteenv{tabular} 

\makeatletter
\let\c@author\relax
\makeatother

\usepackage[
backend      = biber,
natbib       = true,
citestyle    = authoryear,
bibstyle     = authoryear,
date         = year,
url          = true,
urldate      = long,
isbn         = false,
doi          = true,
maxcitenames = 1,
maxbibnames  = 9,
uniquelist   = true
    ]{biblatex}
\renewbibmacro*{doi+eprint+url}{%
    \printfield{doi}%
    \newunit\newblock%
    \iftoggle{bbx:eprint}{%
        \usebibmacro{eprint}%
    }{}%
    \newunit\newblock%
    \iffieldundef{doi}{%
        \usebibmacro{url+urldate}}%
        {}%
    }

\renewbibmacro{in:}{}

\journal{arXiv}

\begin{document}

\begin{frontmatter}



\title{An advanced reliability reserve incentivizes flexibility investments while safeguarding the electricity market}


\author[DIW_KLI]{Franziska Klaucke}
\ead{fklaucke@diw.de}
\author[DIW_KLI]{Karsten Neuhoff}
\ead{kneuhoff@diw.de}
\author[DIW_EVU,KUL]{Alexander Roth}
\ead{aroth@diw.de}
\author[DIW_EVU]{Wolf-Peter Schill\corref{WPS}}
\ead{wschill@diw.de}
\author[DIW_KLI]{Leon Stolle}
\ead{lstolle@diw.de}

\cortext[WPS]{Corresponding author}
\address[DIW_KLI] {DIW Berlin, Climate Policy Department, Mohrenstra{\ss}e 58, 10117 Berlin, Germany}
\address[DIW_EVU] {DIW Berlin, Energy, Transportation, Environment Department, Mohrenstra{\ss}e 58, 10117 Berlin, Germany}
\address[KUL] {KU Leuven, Energy Systems Integration \& Modeling group, Celestijnenlaan 300, 3001 Leuven, Belgium}

\begin{abstract}
To ensure security of supply in the power sector, many countries are already using or discussing the introduction of capacity mechanisms. Two main types of such mechanisms include capacity markets and capacity reserves. Simultaneously, the expansion of variable renewable energy sources increases the need for power sector flexibility, for which there are promising yet often under-utilized options on the demand side. In this paper, we analyze how a centralized capacity market and an advanced reliability reserve with a moderately high activation price affect investments in demand-side flexibility technologies. We do so for a German case study of 2030, using an open-source capacity expansion model and incorporating detailed demand-side flexibility potentials across industry, process heat, and district heating. We show that a centralized capacity market effectively caps peak prices in the wholesale electricity market and thus reduces incentives for investments in demand-side flexibility options. The advanced reliability reserve induces substantially higher flexibility investments while leading to similar overall electricity supply costs and ensuring a similar level of security of supply. The advanced reliability reserve could thus create a learning environment for flexibility technologies to support the transition to climate neutral energy systems. Additionally, an advanced reliability reserve could be introduced faster and is more flexible than a centralized capacity market. While concrete design parameters are yet to be specified, we argue that policymakers should consider the reliability reserve concept in upcoming decision on capacity mechanisms in Germany and beyond. 
\end{abstract}





\begin{acronym}
    \acro{ERAA}{European Resource Adequacy Assessment}
    \acro{CBT}{Constantly Below Threshold}
    \acro{CSP}{Centralized Solar Power}
    \acro{ENTSO-e}{European Network of Transmission System Operators for Electricity}
    \acro{LWP}{Low-wind-periods}
    \acro{MBT}{Mean Below Threshold}
    \acro{NTC}{Net Transfer Capacity}
    \acro{PV}{Photovoltaics}
    \acro{RoR}{Run-of-River}
    \acro{TYNDP}{Ten Year Network Development Plan}
    \acro{VRE}{variable renewable energy}
    \acro{VMBT}{variable-duration mean below threshold}
\end{acronym}

\end{frontmatter}

\input{01_Introduction}

\input{02_Methods}

\input{03_Results}

\input{04_Discussion}

\input{05_Conclusion}

\section*{Acknowledgments}
We gratefully acknowledge internal funding by the German Institute for Economic Research (DIW Berlin) via the Bridge Project ``SichER''. We thank participants of internal DIW Berlin seminars for constructive discussions on previous drafts. This article builds on a longer DIW Berlin project report that includes additional qualitative discussion of different capacity mechanisms and preliminary model results \citep{neuhoff2024security}.

\section*{Code and data availability}
The model code and all input data are available in a public repository: \url{https://gitlab.com/diw-evu/projects/advanced-reliability-reserve}. Part of the input data, all output data and all figures are available in a public repository: \url{https://doi.org/10.5281/zenodo.15662854}.

\section*{Author contributions}

\textbf{Franziska Klaucke:} Investigation, Data curation, Formal analysis, Methodology, Writing - Original Draft
\textbf{Karsten Neuhoff:} Conceptualization, Methodology, Investigation, Writing - Review \& Editing, Funding acquisition
\textbf{Alexander Roth:} Methodology, Software, Formal analysis, Investigation, Data curation, Writing - Original Draft, Visualization
\textbf{Wolf-Peter Schill:} Conceptualization, Methodology, Formal analysis, Investigation, Writing - Original Draft, Funding acquisition
\textbf{Leon Stolle:} Methodology, Investigation, Writing - Original Draft


\newpage
\printbibliography

\newpage
\input{XX_Appendix}

\end{document}

\endinput

%% file: 01_Introduction.tex
\section{Introduction}

The decarbonization of the German energy system largely relies on the use of renewable energy sources \citep{hansen2019full,froemel2025langfristszenarien,luderer2025szenarien}. While wind and solar photovoltaics (PV) provide emission-free electricity, their variable generation patterns present a challenge to the electricity system, in which supply and demand must match at any time. With rising shares of variable renewables, flexibility in the electricity system becomes increasingly important to cope with resulting imbalances \citep{kondziella2016flexibility}. The necessary flexibility can be provided by both the supply and the demand side. Examples include dispatchable power plants, load shifting by industrial consumers and households, electricity storage, heat storage, or intermediary product storage, as well as cross-border electricity trade \citep{Gils.2014, schill2020electricity, roth_geographical_2023, Arnold.2019}. 

Among various flexibility technologies, demand-side options have emerged as promising, yet often under-utilized \citep{leinauer2022obstacles}. By shifting load from hours with little renewable generation to such with high renewable availability, electricity costs of consumers can be reduced. If load shifting lowers the need for expensive peak capacity, it may also incur overall system cost savings \citep{Gils.2014}. Further potential advantages include reduced use of fossil fuels, respective energy imports and carbon emissions \citep{golmohamadi2022demand}, and a potentially lower need for grid expansion \citep{kim2011common}.

Germany currently has an energy only market, which is complemented with different types of reserves to guarantee security of supply in exceptional situations when the wholesale market does not clear \citep{neuhoff2024security}. This particularly includes a capacity reserve, which has never been activated so far \citep{netztransparenz_kapazitätsreserve}. Yet, the growing importance of renewable energy sources and therefore the increased variability of electricity supply has recently reignited a policy debate on security of supply. The previous federal German government (2021~--~2025) considered different policy options for capacity mechanisms \citep{bmwk2024strommarktdesign}, and the current government plans to follow up on this \citep{koalitionsvertrag2025}. The discussion in Germany focuses on the need for and the design of capacity mechanisms, which remunerate the provision of generation capacity to ensure sufficient dispatchable capacity is available when needed \citep{cramton_capacity_2013}. While traditionally designed to support conventional thermal power plants, capacity mechanisms must now be reassessed in light of their potential interactions with demand-side flexibility, which has often been neglected in the literature. In this paper, we analyze how two main types of capacity mechanisms, capacity markets and reserves, interact with demand-side flexibility technologies. To do so, we use an open-source capacity expansion model of the German electricity sector that includes various demand-side flexibility options.

Under textbook assumptions, prices in a perfect energy only market set adequate incentives not only for efficient dispatch of all generation and flexibility technologies, but also for sufficient capacity investments to ensure resource adequacy \citep{joskow_reliability_2007}. However, in reality there are several barriers that prevent sufficient investment, especially in technologies that serve peak loads, which may only run for a few hours per year and thus rely on high scarcity prices to refinance their investment costs. First, investors in such power plants may be unable to recover their fixed and investment costs due to price caps set by the regulator to protect consumers from scarcity prices. These would decrease the revenues of electricity generators, leading to a ``missing money'' problem \citep{joskow_capacity_2008}. Second, even if there was sufficient revenue, investors might not be able to finance their investment due to incomplete risk trading, or ``missing markets'' for long-term risk allocation \citep{newbery_missing_2016, mays2022private}. Third, politically induced rising shares of variable renewable energy generation lead to increasing wholesale price uncertainty. This may exacerbate concerns about insufficient investment in firm capacities and security of supply \citep{jimenez_can_2024}.

As a solution to these challenges, different designs of capacity mechanisms have been discussed and implemented in various electricity markets around the world \citep{bublitz_survey_2019}. Their goal is to provide sufficient long-term investment signals by replacing uncertain revenues from scarcity pricing with constant capacity payments \citep{cramton_capacity_2013}. Two design options that have been widely implemented so far are centralized capacity markets and capacity reserves. There are also less common capacity mechanisms, such as France's decentralized capacity market \citep{european_commission_2016}, from which we abstract in the following.

In a centralized capacity market, the regulator determines the necessary capacity to ensure a desired level of security of supply and contracts this capacity via auctions. The winning bidders then receive a constant capacity payment in addition to the revenues they generate from selling electricity on the wholesale market \citep{kozlova2022combining}. The concept of a capacity reserve, also referred to as strategic reserve, is fundamentally different. Here, the regulator contracts generation capacity outside the wholesale market and exclusively maintains it in a reserve. This reserve capacity is allowed to produce only when a dispatch criterion is reached, for example when the market does not clear or when a certain price level, typically a very high one, is exceeded \citep{bhagwat_effectiveness_2016}. In the following, we focus on a particular type of such a capacity reserve with a moderately high activation price, which we refer to as an ``advanced reliability reserve'' \citep{neuhoff_coordinated_2016}.

The verdict on which type of capacity mechanism is preferable, however, is far from conclusive. Employing an agent-based model calibrated to the German electricity market, \citet{keles_analysis_2016} find that both a centralized capacity market and a capacity reserve can provide security of supply. Despite the capacity market being prone to overcapacity, it is found to be the cheaper design option in the long term. \citet{hoschle_electricity_2017}, modeling capacity expansion as a non-cooperative game for the Belgian market, find that both mechanisms reduce total consumer costs compared to an energy only market. The centralized capacity market is economically more efficient, as capacity in the reserve cannot participate in the balancing market, which is also needed for security of supply. \citet{weiss_market_2017}, comparing market designs in a 100\% renewable Israeli electricity system, conversely find that both a reserve and capacity market increase security of supply, but that the capacity market does so at higher costs due to overcapacity. \citet{jimenez2025capacity} find that capacity reserves, centralized and decentralized capacity markets can improve security of supply compared to an energy-only market, in a decarbonized electricity market of the Netherlands. Capacity markets do so at lower costs while also reducing price volatility relative to the capacity reserve.

Previous analyses, however, have mostly not explicitly addressed the interactions between different types of capacity mechanisms and demand-side flexibility. Potential reasons for this may be that the availability of techno-economic data on flexibility options is limited \citep{fraunholz_role_2021}, and that the aspect of flexibility only had limited relevance in legacy power systems dominated by dispatchable thermal generators. \citet{keles_analysis_2016} distinguish between exogenous scenarios with high or low availability of sheddable load and apply it to an energy only market, but not to capacity mechanisms. \citet{weiss_market_2017} restrict demand-side flexibility to battery storage, which they assume to be the only technology contracted in the strategic reserve. \citet{jimenez2025capacity} consider different consumer groups with different values of lost load able to provide demand response, as well as shedding of industrial heat load. They find that a strategic reserves activates more demand response, as the reserve plants are only activated at an activation price of~\SI{4000}{EUR \per MWh_{el}}, which lies above the costs of demand response. However, the authors do not consider investment costs for demand-side flexibility. \citet{rious_which_2015} calculate potential revenues of demand response operators in France, but do not model the market equilibrium.

Demand-side flexibility can also be considered as a substitute for thermal peak generation capacity \citep{rious_which_2015, keles_analysis_2016, khan_how_2018} and can thus contribute to security of supply \citep{sperstad_impact_2020}. Capacity markets, however, discriminate against demand-side flexibility via multiple channels. Firstly, pre-qualification procedures for capacity markets disadvantage flexibility technologies \citep{cabot_demand-side_2024}, excluding them from obtaining capacity payments through capacity mechanisms \citep{askeland_equilibrium_2019}. For example, different types of storage can be operated for different durations before they run empty, depending on their storage capacity and the filling level. As a consequence, complex de-rating calculations are required to make their contribution to security of supply comparable to firm generation capacity \citep{askeland_equilibrium_2019}. Secondly, capacity markets distort investment decisions by creating a bias for peak generation technologies with low fixed costs but high operating costs \citep{fraunholz_role_2021}. This occurs as technologies with high operating costs refinance a larger share of their costs through scarcity prices. In a capacity market, this revenue stream is replaced by capacity payments and spot market prices are less volatile \citep{mays_asymmetric_2019}. As, however, exploiting this price volatility is the main source of revenue for flexibility technologies, a capacity market reduces their profitability \citep{hogan_follow_2017}. Thirdly, capacity markets distort bilateral contracting and risk trading \citep{neuhoff_coordinated_2016}.

In this paper, we first gather the potentials and techno-economic parameters for providing demand response with various industrial processes and district heating in Germany. We then include this portfolio of demand-side flexibility options in a capacity expansion model of the German electricity market. We apply the model to a 2030 scenario with a high share of renewable generation, comparing a centralized capacity market and an advanced reliability reserve which is activated if wholesale electricity prices reach ~\SI{500}{euro \per MWh_{el}}. We optimize both the investments in different types of demand-side flexibility and their hourly utilization. 

Our contribution to the literature is thus twofold: First, we evaluate the effects of capacity mechanism design on investment and utilization of demand response and storage technologies, based on a comprehensive review of their potential and costs in Germany. Second, we aim to advance the literature on the design of capacity reserves and add to the comparative analysis of different capacity mechanisms.

%% file: 02_Methods.tex
\section{Methods}
\label{sec:method}

\subsection{Capacity expansion model}

To model the German electricity sector, we use the open-source capacity expansion model DIETER \citep{zerrahn_long-run_2017, gaete_dieterpy_2021}. Different versions of this model have been used in previous research to investigate the energy system implications of electricity storage \citep{schill_long-run_2018}, electric heating \citep{schill_flexible_2020,roth_power_2024}, electric mobility \citep{gaetemorales_power_2024,gueret_impacts_2024}, green hydrogen \citep{kirchem_power_2023} and the European interconnection \citep{roth_geographical_2023} in the renewable energy transition. It is a linear program that minimizes overall system costs, i.e., investments and hourly use of various generation and storage technologies, for given, price-inelastic demand and weather data. Its results can be interpreted as the outcome of an ideal, frictionless market with perfect foresight. The model takes all consecutive hours of a full year into account to properly represent renewable energy variability and the use of flexibility technologies.

In this analysis, we use a reduced version of the model that focuses on Germany only, but includes a wide range of demand-side flexibility technology options. Details regarding the modeling, technical specifications, and costs of these demand-side flexibility technologies are provided in the next section. Other key input data for the model include time series of the demand for electricity, heat and green hydrogen, time series on the availability of variable renewables, as well as cost assumptions and bounds for investments in various technologies. While the capacities of variable wind and solar, hydropower and bio energy are fixed exogenously, the remaining part of the power plant fleet is determined endogenously for a challenging weather year with a high peak residual load (2009). This includes in particular the capacities of gas-fired power plants, batteries, hydrogen storage as well as investments in various flexibility options on the demand side. These capacities are then fixed for the model runs of other weather years.

\subsection{Key input data}

Variable renewable energy capacities are fixed at the values targeted by the German government for 2030: Photovoltaics (\SI{215}{GW_{el}}), onshore wind power (\SI{115}{GW_{el}}), and offshore wind power (\SI{30}{GW_{el}}). The capacities of oil-fired, bio energy, hydropower and pumped hydro storage power plants are also fixed (Table~\ref{tab:dieter}). We assume that the coal phase-out will be completed by 2030. This means that neither lignite nor hard coal power plants are available to the model.

In contrast, the model endogenously determines installed capacity of two types of gas-fired power plants (gas turbine, combined-cycle gas), as well as battery and hydrogen storage, including electrolyzers and hydrogen power plants. In addition, the model can choose to invest in various demand-side flexibility options, as discussed in section~\ref{sec:flex_options}. Techno-economic assumptions are summarized in Table~\ref{tab:dieter}. We further assume a CO$_2$ price of~\SI{130}{euro \per ton}.

The annual electric load, excluding the electricity demand for district heating, is assumed to be around \SI{670}{TWh_{el}} for all years. The annual difference between this number and the current demand, which accounts for additional sector coupling, is distributed equally over all hours. This approach neither results in load peaks that are overly inflated, nor in underestimating future demand peaks by implicitly assuming too much flexibility, when compared to more extreme alternative methods such as linear scaling or off-peak distribution of additional demand. We add the amount of electricity required to cover district heating, which ranges between~87 and \SI{103}{TWh_{th}} (average \SI{95}{TWh_{th}}) across all scenarios and years, which corresponds to around 30\% of the space and water heating demand of multifamily and commercial buildings. This translates to roughly 19 to \SI{25}{TWh_{el}} (average \SI{22}{TWh_{el}}) of electric energy. Therefore, total annual electricity demand is around \SI{700}{TWh_{el}}, which ranges between current policy goals and recent system studies \citep{froemel2025langfristszenarien, luderer2025szenarien, aurora2025systemkostenreduzierter}. 

The time series for electricity demand \citep{entso_eraa_2023_demand} and renewable availability \citep{entso_eraa_2023_pecd} come from ENTSO-E's \ac{ERAA} 2023. The time series for heat demand and efficiency of heat pumps are based on \citet{ruhnau_when2heat_2023}. Hence, depending on the weather year, not only the electricity generation of wind and solar PV varies, but also electricity demand related to space heating. 

\subsection{Considered demand-side flexibility options}
\label{sec:flex_options}

We consider demand-side flexibility options (also referred to as demand response) for energy-intensive industries, process heat, and district heating. Technical potentials, parameters, and costs are projected for the year 2030 based on literature data. This includes the potentials for load change, their maximum duration of load change in hours, and associated costs (for a definition see \citet{Hoffmann.2021}).

We assume that the various options only require investments in energy storage or intermediate product storage to unlock demand-side flexibility, while the necessary production capacity already exists (see~\ref{app:DR-Ind}). Note that both the product storage and thermal storage options considered here have similar effects on the power sector as batteries or other electricity storage technologies. Accordingly, their representation in the model is also similar to electricity storage. The details of the calculations, assumptions, and literature data used to determine the demand response potentials are explained in \ref{app:flex_opt}.

\subsubsection{Energy-intensive industry}

We analyze the flexibility potentials of electric steel, aluminum, paper, cement, chlor-alkali process, and air separation separately, relying on the existing literature (see Table~\ref{tab:DR-Ind-Data}). These processes are of particular interest due to their high electricity intensity \citep{Arnold.2018}. 

\begin{figure}
    \centering
    \includegraphics[width=\linewidth]{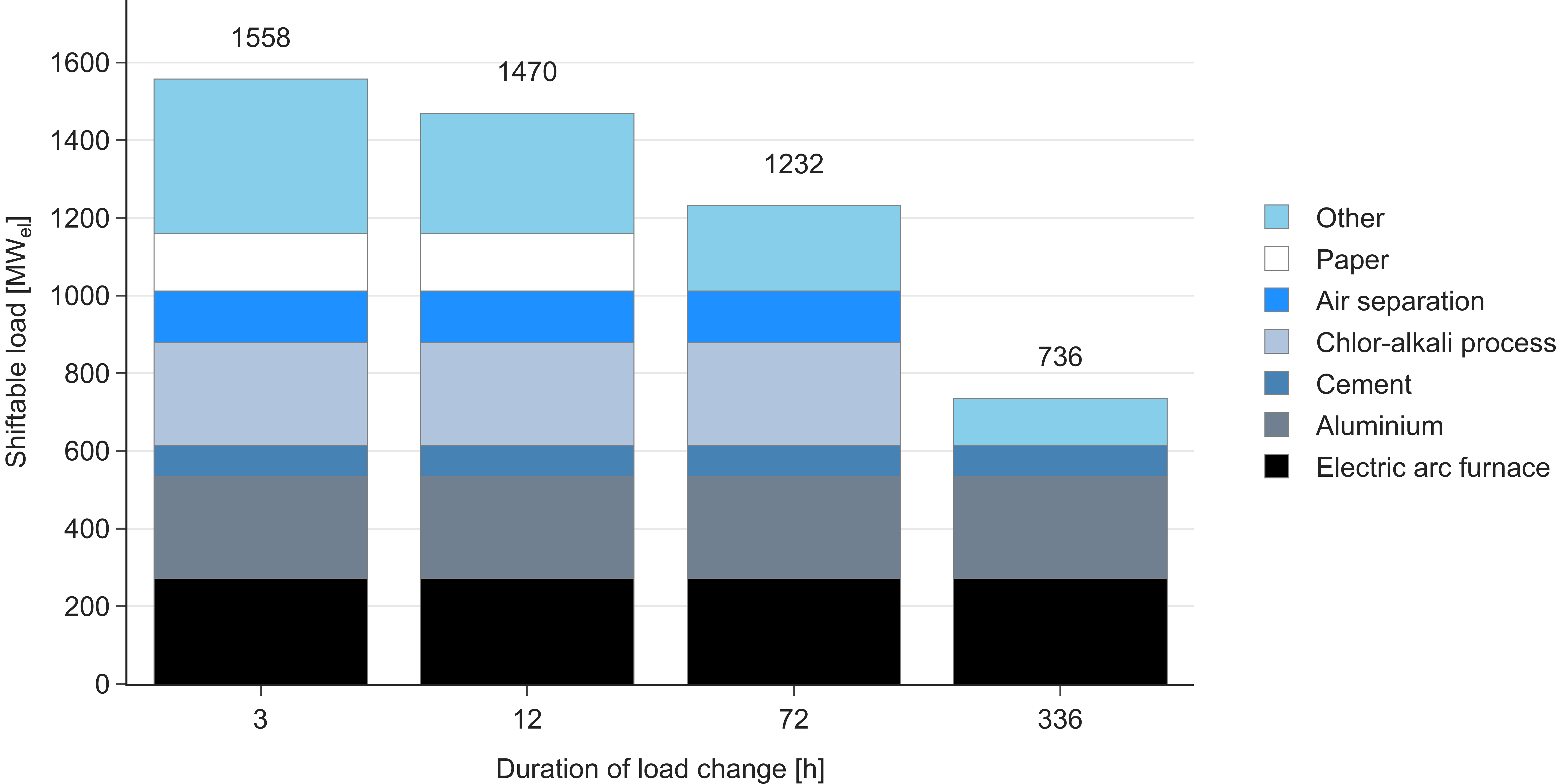}
    \caption{Projected potentials of flexible load of industrial energy-intensive processes for the year 2030 per available duration of load change}
    \label{fig:DR-Ind}
\end{figure}

The flexibility potentials of installed electrical load of other energy-intensive industries are estimated as an aggregate, based on their reported electricity demand \citep{BAFA.2022}. Due to the increasing future electrification, the electricity demand has been scaled up by~\SI{39}{\percent} in 2030 \citep{BostonConsultingGroup.2021}. We assume that~\SI{10}{\percent} of the installed electrical load could be run flexibly (see assumptions in \ref{app:flex_opt}). To reduce complexity, we further assume that electricity-intensive parts of a production process can be shifted in time by storing the products in a physical storage (load shifting). Load shedding, i.e., the temporary reduction of electricity-based production without compensation at a later point, also referred to as ``demand destruction'', is not considered. Figure~\ref{fig:DR-Ind} shows the resulting potentials for flexible loads in energy-intensive industries for~2030. The costs of demand response account for both the costs of retrieving the demand response potential and the investment costs of the storage system. The applied costs and assumptions are described in the~\ref{app:DR-Ind}.

The model decides endogenously, based on storage and activation costs, which amount of energy can be shifted at maximum. For example, all technologies with a load change duration of three hours (see \ref{app:DR-Ind} assumption~6) and an overall shiftable load of~\SI{1558}{MW_{el}} (see figure~\ref{fig:DR-Ind}) have the potential to add a maximum storage size up to~\SI{4674}{MWh_{el}}. Technologies with correspondingly higher durations can also be equipped with larger storage. 

\subsubsection{Process heat}

Process heat constitutes the largest share of industrial energy consumption in Europe. Currently, natural gas is the dominant energy carrier, supplying~\SI{35}{\percent} of total energy demand of process heating in 2019 in the EU-27 \citep{FraunhoferISI.2024}. The electrification of process heat is therefore crucial to achieve future climate targets \citep{BostonConsultingGroup.2021}. In contrast to the seasonal demand profile of space heating, the demand for process heat does not vary throughout the year, as we assume no seasonality of industrial production \citep{Gils.2014}. 

We consider process heat up to a temperature level of~\SI{500}{\degreeCelsius}, at which heat pumps and resistance heaters (electrode boiler) can be used on an industrial scale according to the current state of the technology \citep{Thiel.2021,FraunhoferISI.2024}. Thereby, the process heat is provided by high-temperature heat pumps for temperatures up to~\SI{160}{\degreeCelsius} and by resistance heaters for temperatures between~\SI{200}{\degreeCelsius} and~~\SI{500}{\degreeCelsius} \citep{Fleiter.2023}. We base the calculation of future electrical energy demand for the provision of process heat on \citet{Kemmler.2017}. The metal and mineral industry is excluded because their production processes mainly require heat directly in the processes and not indirectly, e.g., via process steam \citep{Fleiter.2023}. Additionally, the required temperatures are usually at the upper limit of the temperature levels considered here \citep{Kemmler.2017}. We assume a standard packed-bed thermal energy storage system with a storage efficiency of~\SI{90}{\percent} \citep{Profaiser.2022b} and with a maximum installed storage capacity of~\SI{30}{\percent} of the considered installed electrical capacity for process heat by the year 2030. The maximum thermal storage period is 72~hours. Due to low investment costs, the installed excess capacity for process heat is realized by resistance heaters with a thermal efficiency of~\SI{99}{\percent} \citep{Fleiter.2023}. Storage losses as a percentage of capacity per day are~\SI{3}{\percent} \citep{Arnold.2019}.

Table~\ref{tab:prozessheat} summarizes the data for process heat. The detailed calculations, data basis and assumptions are described in \ref{app:process_heat}. The power rating of process heat installations is fixed in the model. However, the amount of energy, i.e., the size of a thermal storage, is endogenous.

\begin{table}[h]
\footnotesize
\caption{Projections of future energy demand for the provision of industrial process heat}
\label{tab:prozessheat}
\begin{threeparttable}
\begin{tabulary}{\textwidth}{L|L|L|L}
\toprule
Temperature range process heat [\si{\degreeCelsius}] & 
Thermal energy demand for process heat\tnote{a} [\si{TWh_{th} \per a}] & 
Electrical energy demand for high tem\-pe\-ra\-ture heat pumps\tnote{b} [\si{TWh_{el}}] & 
Electrical energy demand for resistance heaters\tnote{c} [\si{TWh_{el} \per a}] \\
  \midrule
    $ < 100 $        &  67     & 18.1 &                    \\
    100 -- 160          &  70.2   & 19   &                    \\ 
    160 -- 500          &  33.1   &      & 33.4               \\
    \textbf{Sum}     &  170.2  & \multicolumn{2}{c}{70.5}  \\
   \bottomrule
\end{tabulary}
\begin{tablenotes}[para,flushleft]\footnotesize \item[a] \citet{Kemmler.2017}; \item[b] Assumed coefficient of performance (COP) high temperature heat-pump: 3.7 \citep{Fleiter.2023}; \item[c] Assumed efficiency factor resistance heater~\SI{99}{\percent} \citep{Fleiter.2023}.
\end{tablenotes}
\end{threeparttable}
\end{table}


\subsubsection{District heating}

To consider the flexibility potential of thermal storage in heating networks, we include a simplified district heating market in the model. We assume that an average of~\SI{95}{TWh_{th}} of heat per year is provided by district heating, which corresponds to~\SI{30}{\percent} of the German heat demand in residential and commercial buildings. In the model, the entire district heating demand is met by large-scale ground-sourced heat pumps. Load shifting is enabled by sensible thermal storage systems using water as a storage medium. The storage efficiency is~\SI{90}{\percent} with standing losses of~\SI{2}{\percent} per day \citep{Arnold.2019}. The size of the thermal storage tank is endogenously determined by the model, while the heat and electricity outputs of heat pumps are exogenously given. Investment costs for the thermal storage amount to~\SI{50}{EUR \per kWh_{th}} \citep{Arnold.2019}. Decentralized heat pumps and their buffer storage capacities are not modeled. Despite typically small storage sizes, they may also contribute some short-duration flexibility to the electricity system \citep{roth_power_2024}.

\subsection{Two policy scenarios}\label{subsec:scenarios}

We investigate two policy scenarios with different capacity mechanisms: a centralized capacity market and an advanced reliability reserve.

The centralized capacity market supplements the electricity market by specifying a desired amount of firm capacity in the model. We require that \SI{101.3}{GW_{el}} of capacity are provided, which can be covered by natural gas (open or combined-cycle gas turbines), oil, bio energy, hydro reservoirs, and long-duration hydrogen storage. The capacity target is set so that in all modeled weather years the residual load (electric load excluding electricity demand for heat minus generation from variable renewables) can be covered by the generation capacity contracted in the capacity market, including a safety margin of five~\si{GW_{el}}. However, as the capacities of the other technologies are fixed, only gas and hydrogen power plants are available to the model as free variables.

In the scenario with an advanced reliability reserve, the electricity market is complemented by reserve power plants located outside the electricity market. These run only when the electricity price reaches the defined activation price of~\SI{500}{euro \per MWh_{el}}. We choose this price level to satisfy and balance two objectives. First, the price must be high enough to provide sufficient investment incentives into new generation and flexibility capacity in the wholesale market. Secondly, it must be low enough to be credible: if the activation price is too high, the regulator may face political or societal pressure to activate the reserve already at lower prices to shield consumers from high energy costs (more discussion on this in section~\ref{sec:discussion}). The size of the reserve is determined in such a way that the sum of firm capacity in the wholesale market and in the reserve equals the size of the capacity market (\SI{101.3}{GW_{el}}), resulting in a similar level of resource adequacy.

To properly account for temporal correlations between renewable energy scarcity and high demand peaks in winter periods, the model is run in a ``summer-to-summer'' mode (i.e., from July to June). When, for instance, the year ``2009'' is mentioned, we refer to the period ``July~2009~--~June~2010''.

For each scenario, we apply a two-step procedure: In the first step, we obtain a least-cost solution and optimal capacities of power plants and demand flexibility technologies for a single weather year. We use the year 2009, which is the weather year with the highest residual load in our sample, and has already served to determine the size of the capacity market. In a second step, the model is run in a dispatch-only mode, using the optimal capacities of the first step, and is iterated over the seven weather years 2008~--~2014 to analyze the effects on prices and the rest of the electricity sector.

Importantly, we assume that demand-side flexibility technologies can neither participate in the capacity market nor in the reserve. In both scenarios, demand-side flexibility technologies thus do not receive any payments from capacity mechanisms, but have to cover their investment and operating costs exclusively by revenues from the wholesale market.

%% file: 03_Results.tex
\section{Results} \label{sec:results}

\subsection{Overall power plant portfolio}

\begin{figure}[tb]
    \centering
    \includegraphics[width=\textwidth]{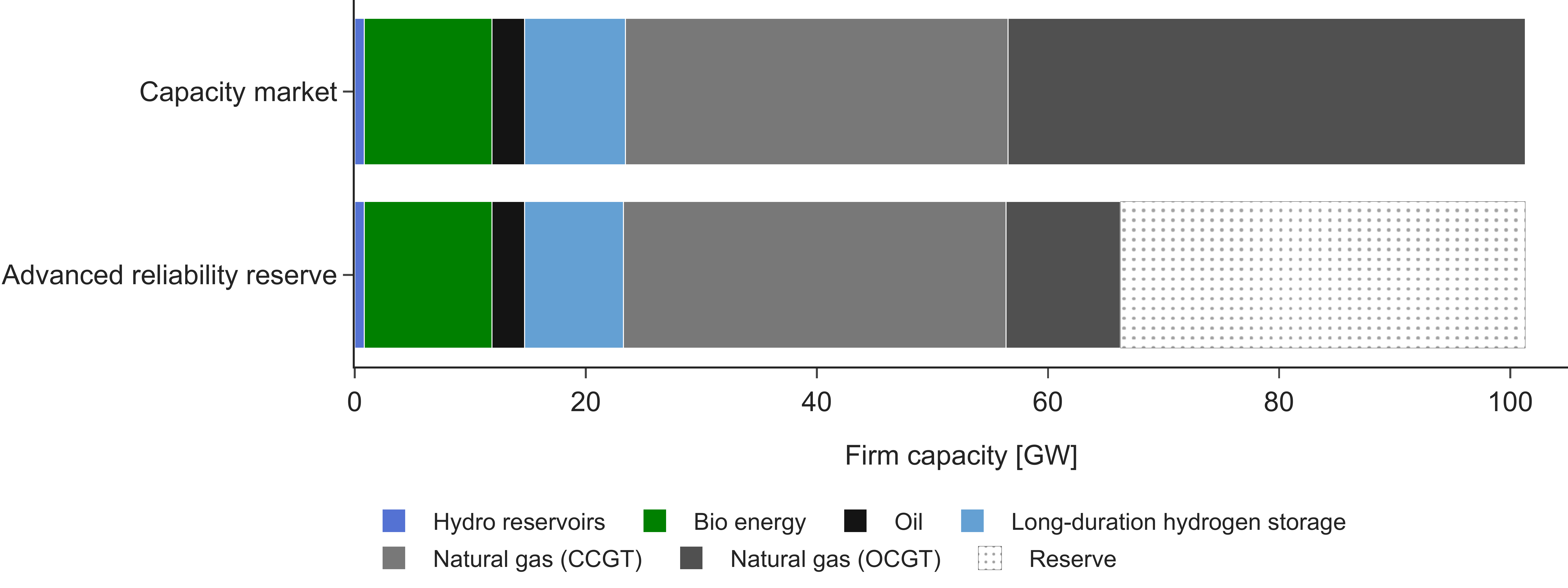}
    \caption{Installed firm capacity by technology for the capacity market and the advanced reliability reserve.}
    \label{fig:firm_capacity}
\end{figure}

The optimal capacity portfolio, resulting from the first step of the optimization, differs between the two capacity mechanisms (Figure~\ref{fig:firm_capacity}). In the advanced reliability reserve, the installed capacity that directly participates in the wholesale market, namely open-cycle gas turbines (OCGTs), is significantly lower than in the centralized capacity market. Instead, extensive generation capacities (\SI{35}{GW_{el}}) are held in the reserve that do not participate in the wholesale market.

While the capacity portfolio of the reliability reserve is not part of the optimization, we assume that it consists of newly built open-cycle gas turbines, which have low specific investment costs. In reality, however, the reserve could also absorb existing plants that leave the wholesale market. This has already been the case for the existing capacity reserve in Germany.

\subsection{Impacts on wholesale electricity prices}

\begin{figure}[!tb]
    \centering
    \includegraphics[width=\textwidth]{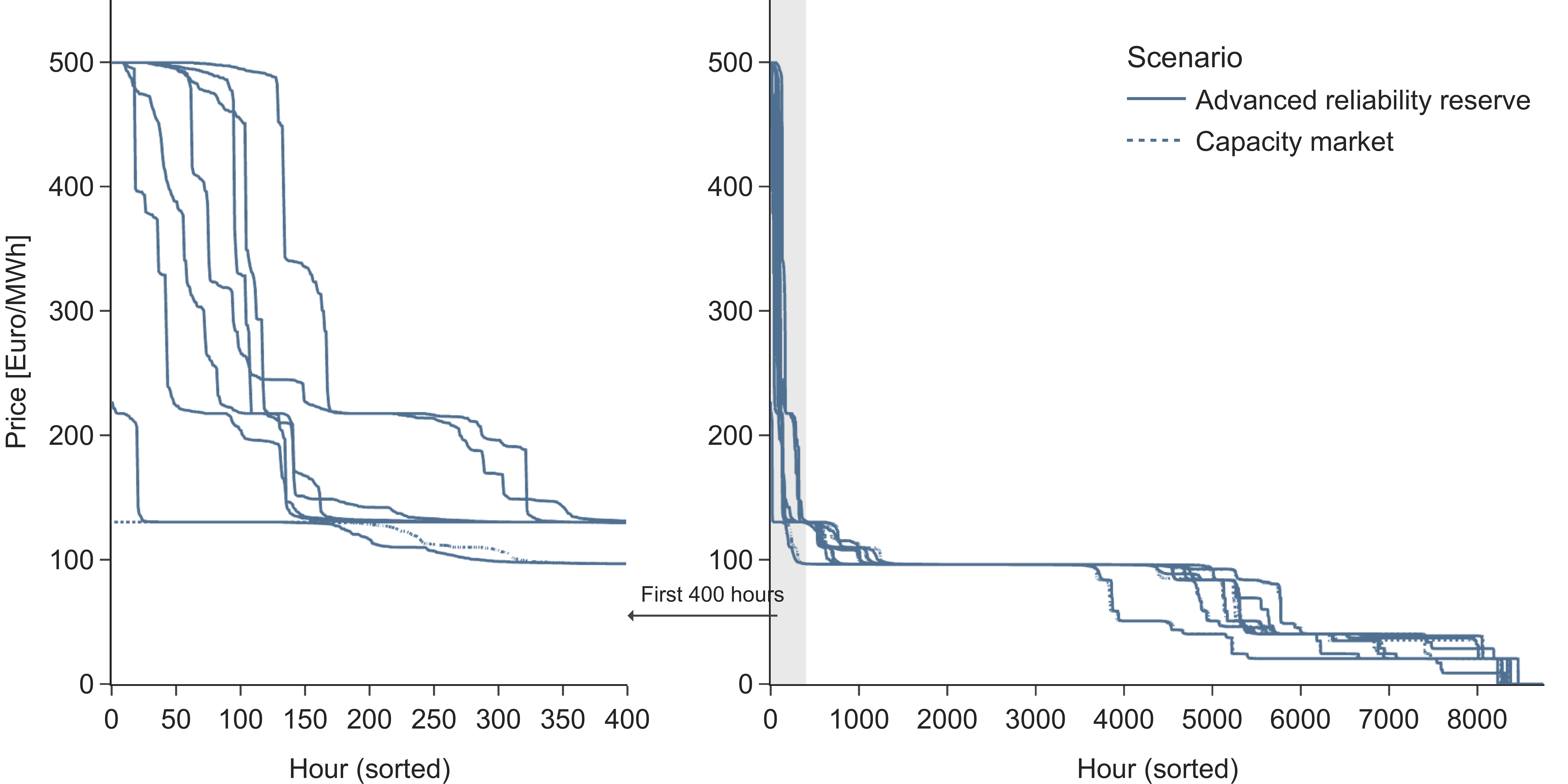}
    \caption{Wholesale market price-duration curves for both capacity mechanism scenarios and different weather years (2008 -- 2014).}
    \label{fig:price_duration}
\end{figure}

The main difference between a centralized capacity market and an advanced reliability reserve is the ability of the capacities covered by these mechanisms to participate in the wholesale electricity market. While power plants that take part in a centralized capacity market can fully operate on the wholesale electricity market, power plants contained in the reserve cannot, but are only activated by the regulator in case the wholesale market price reaches the activation price. This difference has profound consequences for the price formation in the wholesale market. We illustrate this with price-duration curves for several weather years, which sort all hourly prices of a year in descending order, starting with the highest price on the left-hand side (Figure~\ref{fig:price_duration}).

Depending on the weather year, prices differ greatly between a capacity market and a reliability reserve (left panel, Figure~\ref{fig:price_duration}) for a maximum of around 400~hours. In a centralized capacity market, there is always sufficient generation capacity from gas-fired power plants during these hours, and their marginal costs determine the price in all peak hours. The resulting maximum price is around~\SI{130}{euro \per MWh_{el}}, which corresponds to the marginal production costs of a gas turbine in the model, considering fuel and carbon prices as well as other variable costs. 

In contrast, wholesale prices gradually rise to~\SI{500}{euro \per MWh_{el}} in some hours in the scenario with an advanced reliability reserve. The electricity price never exceeds~\SI{500}{euro \per MWh_{el}}, as this price triggers the activation of the reserve, which has always sufficient capacity to serve demand.

During most other hours of the year, the results for the centralized capacity market and the reliability reserve hardly differ (right panel, Figure~\ref{fig:price_duration}). On the right-hand side of the price-duration curve, the differences between weather years turn out to be substantially greater than those between the two policy scenarios.

\subsection{Investments in flexibility technologies}

In line with the different wholesale prices patterns emerging in the two scenarios, potential revenues for flexibility technologies greatly differ. Hence, there are also different incentives for respective investments in the second step of the optimization.

\begin{figure}[!tb]
    \centering
    \includegraphics[width=\textwidth]{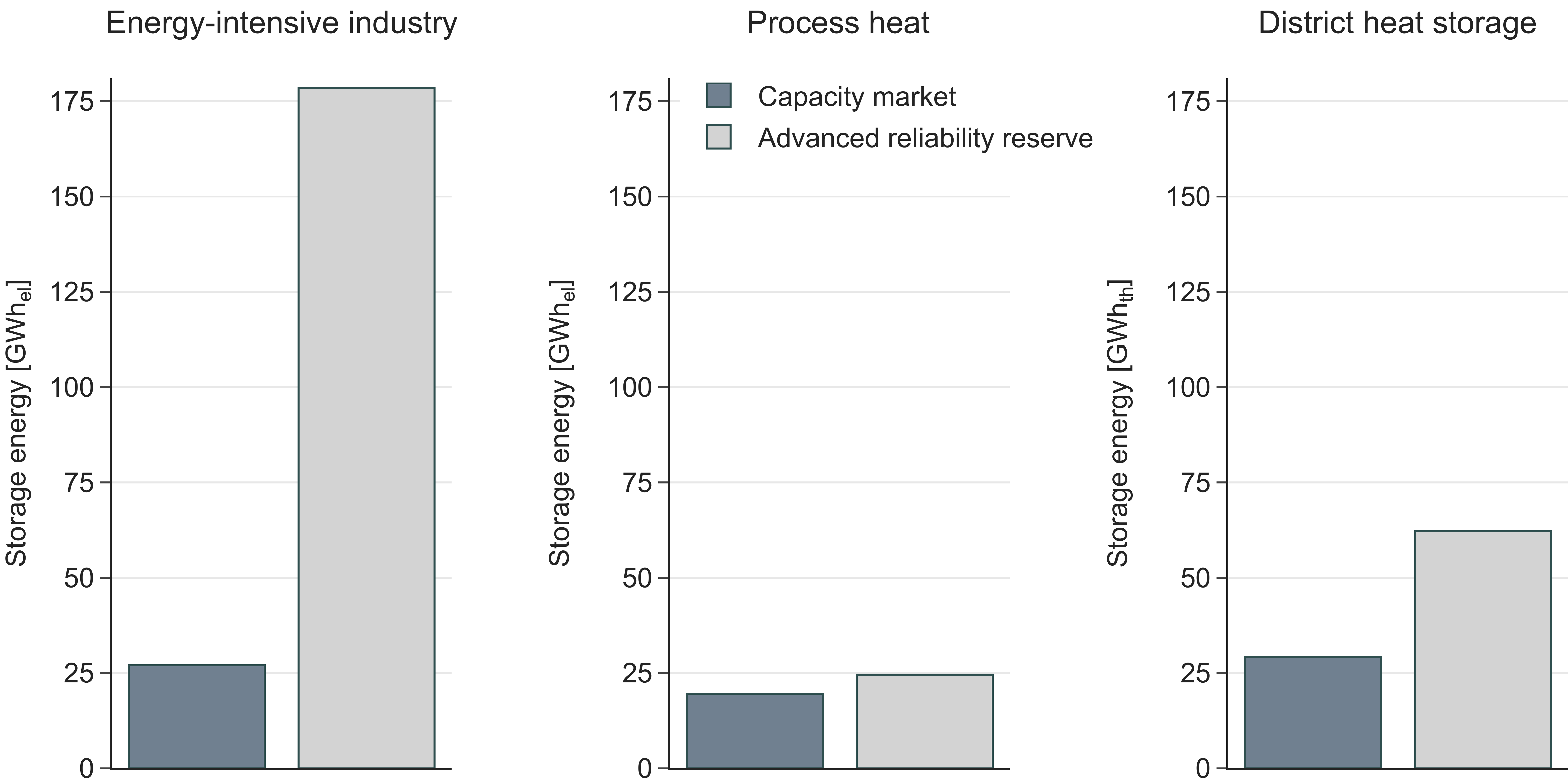}
    \begin{minipage}{\textwidth}
    \end{minipage}
    \caption{Storage capacities for enabling demand-side flexibility for both capacity mechanism scenarios.}
    \label{fig:investments}
\end{figure}

With its flat peak prices, the centralized capacity market leads to substantially lower investments in different demand-side technologies in the electricity market compared to the advanced reliability reserve (Figure~\ref{fig:investments}). The centralized capacity market leads to such a high capacity of firm power plants that prices only rise to the variable costs of the most expensive plant. This reduces potential revenues of flexibility technologies, which are thus crowded out on the wholesale market. 

In contrast, investments into flexibility technologies are much higher if an advanced reliability reserve complements the electricity market. Model results show that investments into storage for flexibilizing processes in the energy-intensive industry are nearly seven times higher than under a capacity market. Investments into process thermal storage increase by around a quarter. In the case of district heating energy storage, the reliability reserve leads to more than twice as much investment. As discussed below, the precise model outcomes hinge on the weather year. However, the qualitative effects remain robust also using different weather years (Figure~\ref{fig:flex_years}).

\subsection{Activation of the advanced reliability reserve}

\begin{figure}[htb]
    \centering
    \includegraphics[width=\textwidth]{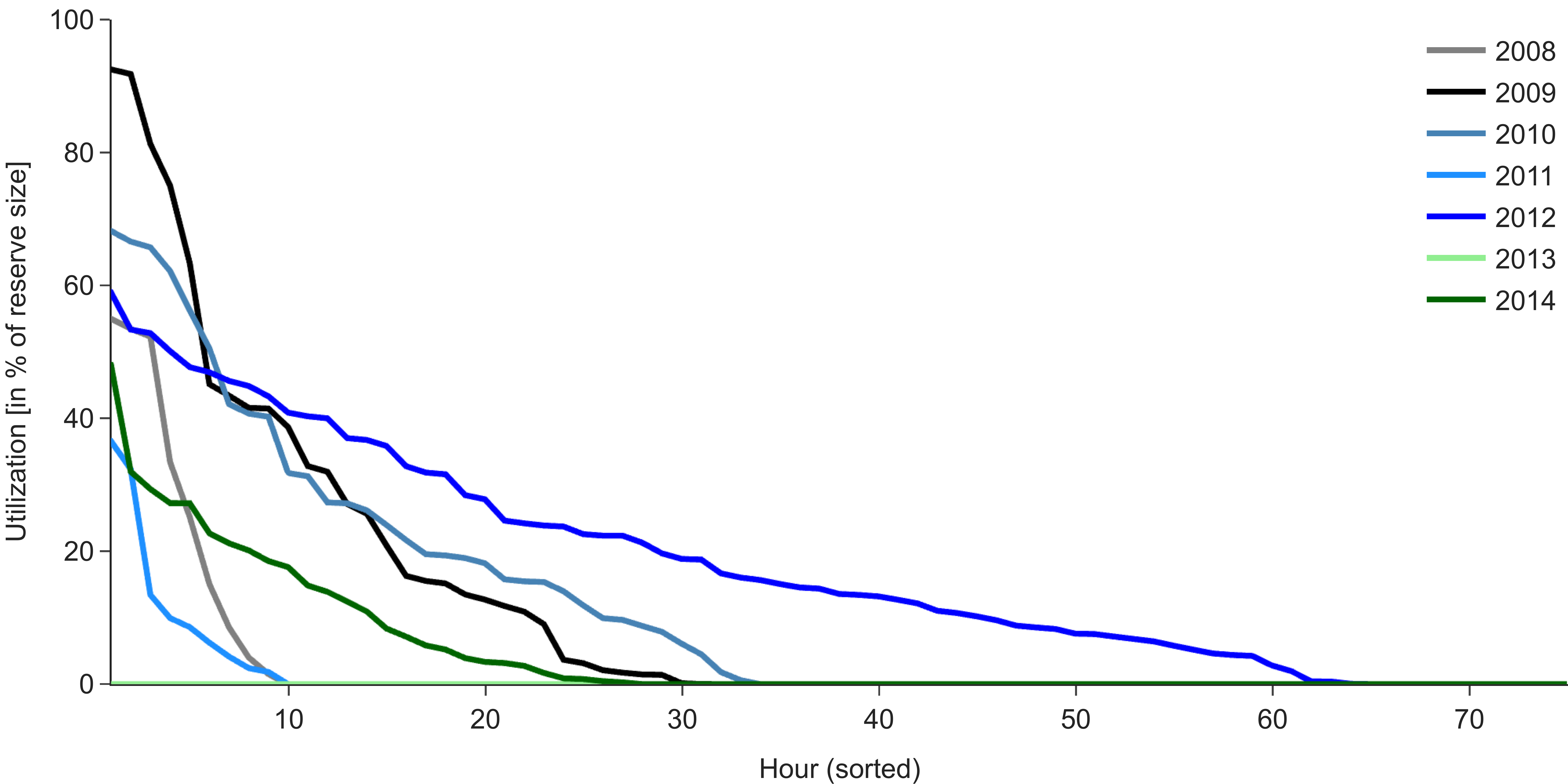}
    \caption{Operating hours of the advanced reliability reserve in different weather years.}
    \label{fig:operating_hours}
\end{figure}

In each year, the reliability reserve is only activated for very few hours. Depending on the weather year, it is used between 0~and~64~hours, as shown in Figure~\ref{fig:operating_hours}. And even when the reserve is activated, it is mostly used at less than half of its capacity. In the seven weather years modeled here, the reserve is activated for only 175~hours (0.29~\% of all hours). 

Considering the extremely rare activation of the reserve in only a few hours with very high output, it seems plausible that the reserve could in reality be smaller than modeled here. A similar reasoning applies for the centralized capacity market. Ultimately, a weather situation with cold temperatures and low wind and solar power generation – as observed on a December day in~2009 – leads to such a high remaining demand for electricity that the reserve must be dimensioned almost twice as large as would be necessary for all other 2554~days covered in the analysis. 

\subsection{Average electricity supply costs}

\begin{figure}[tb]
    \centering
    \includegraphics[width=\textwidth]{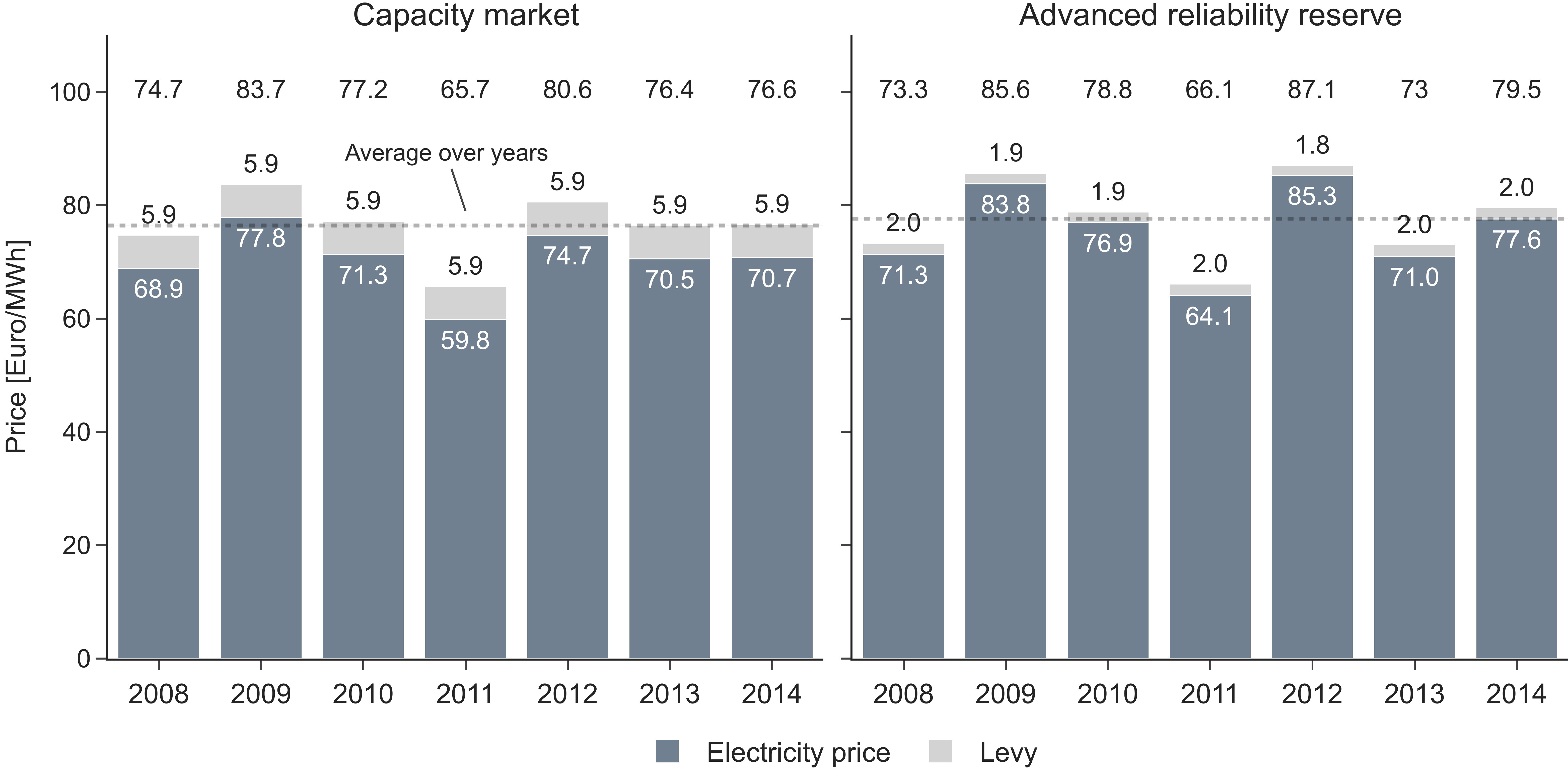}
    \caption{Average wholesale electricity prices and levies for capacity mechanisms in different weather years (2008 -- 2014).}
    \label{fig:average_prices}
\end{figure}

Figure~\ref{fig:average_prices} shows average annual wholesale prices as well as the required payments for the capacity mechanism, allocated to the total electricity demand as a levy per megawatt hour. We define the sum of the two as average supply costs of electricity. Note that we do not consider other price components of end user tariffs here, such as grid fees or taxes. Average supply costs depend primarily on the assumed fuel prices, which we hold constant in our analysis, and should thus not be interpreted as policy-relevant forecasts for~2030.

We first compare capacity payment differences between the two mechanisms. The respective cost allocation for the capacity market is~\SI{5.9}{euro \per MWh_{el}}, and varies between 1.8 and \SI{2.0}{euro \per MWh_{el}} for the reserve. The main reason for the higher levy in the capacity market scenario is that all power plants receive a payment that reflects the annualized investment costs plus annual fixed costs of open-cycle gas turbines, which are the marginal power plants that receive the capacity payment. As the capacity portfolio does not differ between weather years, also this capacity payment is constant across years. In the case of the advanced reliability reserve, the capacity payments only finance the power plants in the reserve. In contrast to the capacity market, the reserve plants also get compensation for their variable costs through the capacity payments, and not only for their capacity costs, when they are activated. The levy is the sum of capacity and activation payments for all plants in the reserve, minus the revenues from selling the respective amount of electricity at the activation price. Accordingly, the levy slightly varies across years and is slightly lower in years in which the reserve is activated more often (e.g., in~2012).

Average wholesale electricity prices are slightly lower in the capacity market scenario than in the scenario with a reliability reserve, as prices in the former are effectively capped at the marginal costs of open-cycle gas plants. Yet, differences between weather years are larger than differences between the two capacity mechanisms.

Combining average wholesale prices and levies, overall electricity supply costs hardly differ between the two capacity mechanisms, averaged across the years~2008 until~2014. In the current model calibration, the capacity market leads to slightly lower average supply costs of \SI{76.4}{euro \per MWh_{el}} than the reliability reserve at~\SI{77.6}{euro \per MWh_{el}}. The economic inefficiencies introduced by both capacity mechanisms largely balance each other in our model parameterization: the capacity market discriminates against flexibility technologies, and the advanced reliability reserves leads by design to prices above the costs of the marginal power plant when it is activated.

%% file: 04_Discussion.tex
\section{Discussion}\label{sec:discussion}

\subsection{Model limitations and their qualitative effects}

The primary objective of this model analysis is not to produce precise forecasts of the composition of the German power plant fleet, future energy prices, or the dimensioning of capacity mechanisms. Rather, our analysis seeks to generate insights into how different capacity mechanism designs influence the investments and utilization of demand-side flexibility. For tractability, the electricity sector model comes with several simplifications.

First, the analysis is based on a deterministic model with perfect foresight. This assumption may lead to an overly optimistic use of demand-side flexibility technologies, as they can ideally adjust to variable electricity generation from wind and solar power and respective wholesale market price fluctuations. However, model distortions related to perfect foresight are likely to be larger for the operation of long-duration storage technologies \citep{schmidt2025on} as compared to shorter-duration flexibility options which are in the focus here.

Second, we do not explicitly model electricity grids, but implicitly assume that sufficient grid capacity is available to avoid any bottlenecks in the transmission and distribution grids. Given that this assumption is unlikely to hold in Germany in the foreseeable future, our analysis does not capture potential interactions of capacity mechanisms or demand-side flexibility options with grid constraints. Consequently, we do not quantify potential benefits of an advanced capacity reserve when it comes to the spatial allocation of dispatchable power plants or other flexible capacities in grid-constrained settings, or reductions of grid-capacity needs.

For the sake of clarity and simplicity, we further exclude other flexibility options such as cross-border electricity exchange or flexible sector coupling technologies like electric vehicles or decentralized heat pumps in households \citep{brown_synergies_2018}. As a result, we underestimate the flexibility potential in the German power sector compared to real-world conditions, and overestimate the reliance on dispatchable capacities in Germany. The outcomes of this study should therefore be interpreted as an upper bound of the thermal power plant portfolio size for the weather years analyzed here. 

In addition, we base our analysis on a sample of seven weather years. While this can be considered an improvement compared to many analyses that use a single weather year only, we may not capture rare but very prolonged renewable energy drought events, which are expected to become more relevant in future renewable energy systems \citep{gotske2024designing, kittel_measuring_2024, kittel2025coping}. Expanding the dataset to include a broader range of weather years that covers more extreme events could lead to higher optimal capacities of firm generation, long-duration storage or demand-side flexibility technologies. 
The capacity results of this analysis are based on the challenging weather year 2009, yet results differ when other weather years are used (see Figure~\ref{fig:flex_years}). However, the qualitative result of higher investments into flexibility with a reliability reserve remains stable across years. 

Finally, our results hinge on the assumption that demand side-flexibility technologies cannot effectively participate in centralized capacity markets. This is in line with literature that concluded that capacity markets typically discriminate against demand-side flexibilities \citep{cabot_demand-side_2024}. It should be noted that some existing capacity markets (for example in Ireland and the UK) do partially allow for the inclusion of demand-side flexibilities. Yet, this would require, for example, applying non-discriminatory de-rating factors \citep{fraunholz_role_2021}, the design of which appears to be a major challenge. 

\subsection{Long-term outlook beyond 2030}

The analysis is conducted for the year 2030, based on policy-relevant assumptions on electricity demand and renewable generation shares. However, this~2030 perspective neglects the fact that the use of variable wind and solar power is planned to increase significantly in Germany in the longer run, which is in line with the country's 2045 climate neutrality target. Accordingly, the system benefits of demand-side flexibility are likely to increase further over time and are thus underestimated in our analysis that implicitly assumes a 2030 long-run equilibrium. 

As the reliability reserve unlocks much more demand-side flexibility, it is likely to become also more advantageous over time compared to a centralized capacity market, which crowds out demand-side flexibility via an oversupply of dispatchable power plants.

Furthermore, the modeled flexibility potentials can be considered a lower bound estimate for a long-run decarbonized energy system beyond the year 2030: for example, the model does not yet consider flexibility potentials in households and commerce, as well as flexibility related to electric vehicles, which are all likely to become more relevant beyond 2030. Significant demand-side flexibility potential can still be tapped here in the future, particularly against the backdrop of advancing electrification and sector coupling. 

\subsection{Advanced reliability reserve design considerations and policy implications}

Beyond the flexibility benefits of the advanced reliability reserve quantified in our model analysis, its practical implementation appears to have further advantages compared to a centralized capacity markets. 

In Germany, an advanced reliability reserve could be implemented quickly. The German regulator already has substantial experience with the tendering and operation of different types of capacity reserves \citep{neuhoff2024security}. In an initial phase, the reliability reserve could even consist of power plants that are already part of existing reserves or that will be decommissioned in the coming years for economic reasons.

A central design characteristic of the reliability reserve is determining its size. While centralized capacity markets are prone to oversizing \citep{weiss_market_2017, jimenez2025capacity}, for example because of high levels of political or regulatory risk aversion, this may also be true for reliability reserves. Yet, the reserve appears to be significantly more adaptable and less irreversible than a centralized capacity market. This is particularly true as decisions on the size of the reserve do not have a direct impact on supply and demand and thus on wholesale market price trends, and ultimately on the incentives for investments in flexibility and generation capacity in the electricity market.  

The adaptable size of the reserve could further emerge as a great advantage when the optimal reserve dimensioning might vary over time because of ongoing technological developments. Three energy market characteristics likely decrease the reserve requirements: first, an increasing price elasticity of electricity demand in addition to the specifically modeled investments in demand-side flexibility could help to reduce the size of the reserve. In the same way, cross-border European electricity trade could help to reduce the reserve capacity, similarly to its mitigating effect on storage requirements \citep{roth_geographical_2023}. Third, there may be further potential to negotiate additional contracts with large consumers to reduce their peak loads for the very few peak hours in which the reserve would be activated.

Another important design parameter of the reliability reserve is its activation price. We propose~\SI{500}{euro \per MWh_{el}} to consider the trade-off between regulatory credibility of maintaining this price, and sufficient investment incentives via wholesale market prices. In other words, the activation price has to be low enough to be credible, and high enough to provide investment incentives.

So far, real-world reserves have often been only activated when the electricity market does not clear, which implies extremely high activation prices. In Germany, the upper reference price limit on the day-ahead market is currently at \SI{4000}{euro \per MWh_{el}} \citep{ACER2023}, which is also used as an activation price for the reserve by \citet{jimenez2025capacity}. However, we argue that such a high activation price might not be credible because the regulator would be pressured to mitigate such high prices, especially when they persist over a longer time, by activating the reserve already at lower prices. For example, the 2025 German coalition agreement includes a statement that existing reserve plants could be dispatched with the aim of lowering wholesale market prices \citep{koalitionsvertrag2025}. We argue that such interventions are unlikely to happen in case of a moderately high activation price of~\SI{500}{euro \per MWh_{el}} and thus consider this price to be ``low enough''. 

At the same time, such an activation price appears to be ``high enough'' to provide sufficient incentives for investments into new generation capacity and demand-side flexibility. We consider it unlikely that investors of power plants or storage options would rely on much higher peak prices than~\SI{500}{euro \per MWh_{el}} in their investment decisions. In contrast, activating the reserve at substantially lower prices, e.g.,~at~\SI{150}{euro \per MWh_{el}} as presented by \citet{steag2025}, may deteriorate investment incentives in the wholesale market.

\subsection{Alternative capacity mechanisms}

Beyond the centralized capacity market and the reliability reserve analyzed here, further capacity mechanisms have been discussed in Germany and beyond. In particular, our analysis abstracts from decentralized capacity market concepts. In theory, decentralized mechanisms such as mandatory hedging on future markets or a decentralized capacity market designs could also provide incentives for the activation of demand-side flexibility options similar to the reliability reserve. However, the practical feasibility and functionality of these decentralized mechanisms with regard to effectively ensuring a high level of security of supply is yet to be demonstrated \citep{neuhoff2024security}. We consider the reliability reserve to be a more viable alternative for practical implementation. Nonetheless, future research may further investigate the relative benefits of the reliability reserve and decentralized mechanisms.

%% file: 05_Conclusion.tex
\section{Conclusion} \label{sec:conclusion}

Both a centralized capacity market and an advanced reliability reserve are generally suitable to ensure security of supply in electricity markets. To reflect the increasing need for flexibility in power markets based on variable wind and solar energy, capacity mechanisms should be designed to provide strong incentives for the development of various flexibility technologies. Our model results illustrate that a centralized capacity market does not deliver these incentives in a 2030 scenario for Germany. Wholesale electricity prices are effectively capped at the variable costs of the marginal power plant, in our case an open-cycle gas turbine. This severely limits the revenue opportunities for demand-side flexibility options that rely on higher peak prices. In comparison, the advanced reliability reserve allows higher wholesale price spreads up to the activation price and thus unlocks significantly more investment into demand-side flexibility.

An advanced reliability reserve, properly designed to guarantee the security of supply while maintaining investment incentives, could thus be a preferable alternative to a centralized capacity market. It avoids very high wholesale prices above the activation price of~\SI{500}{euro \per MWh_{el}}, but still allows for sufficient price variation to enable investments in demand-side flexibility. Overall, the reserve could thus strengthen the wholesale market environment. Furthermore, it creates a learning environment for flexibility technologies, which become ever more relevant with increasing shares of variable renewable energy sources. This appears to be beneficial during the transition to climate neutral energy systems. An advanced reliability reserve could also be introduced faster and is more flexible than a centralized capacity market. We thus conclude that policymakers should consider the reliability reserve concept in upcoming decision on capacity mechanisms in Germany and beyond.

Considering the knowledge gaps that still exist, we see promising avenues for further research. For instance, the interplay of capacity mechanisms with support mechanisms for renewable electricity generation --- which are also about to be reformed in Germany --- are of interest. More detailed analysis of the interactions of capacity mechanisms with decentralized options for flexible sector coupling could also be explored, including prosumers who operate batteries, electric vehicles or heat pumps. Last, but not least, further research into the European coordination of the design, procurement, and operation of capacity mechanisms would be desirable.

%% file: XX_Appendix.tex
\appendix

\section{Input data for the capacity expansion model}

\input{Tables/capacities}

\newpage

\section{Assumptions and database for demand-side flexibility options}
\label{app:flex_opt}

\subsection{Demand response potential in energy-intensive industry}
\label{app:DR-Ind}
The following assumptions were taken as a basis for forecasting the future flexibility potential of energy-intensive industry:
\begin{enumerate}
   
    \item Demand response in the form of load relinquishment is considered for the electric steel and aluminum industries due to their technical process conditions; load shifting is assumed for all other processes due to the high costs of lost production.
     \item When differing load shifting potentials were reported in the literature, the maximum technically feasible potential has been used.
    \item If no other information is available, it is assumed that the load flexibility potential of a process is~\SI{10}{\percent} of its installed capacity.
    \item  The installed electrical load of non-specific energy-intensive industries is based on the date of \cite{BAFA.2022}. The values have been adjusted by the considered electrical load of the separately analyzed industries and those industries that are not suitable for demand response (e.g. electrified public transport).
    \item According to the study of \cite{BostonConsultingGroup.2021}, the electricity demand of the non-specific energy-intensive industries will increase by~\SI{39}{\percent} until 2030 due to the increasing electrification of industry.
    \item Due to the lack of detailed data in the literature on the duration of load change, it is assumed that the load flexibility potentials are available proportionally for~\SI{3}{\hour}, \SI{12}{\hour}, \SI{72}{\hour} and~\SI{336}{\hour}. The maximum available duration of load change has been determined according to their technical requirements.
   \end{enumerate}
The basis of calculations, including references, is shown in table~\ref{tab:DR-Ind-Data}.
\begin{center}
\begin{table}[!ht]
\footnotesize
\begin{threeparttable}
\caption{Data and references on the demand response potentials of energy-intensive industrial processes}
\begin{tabularx}{\linewidth}{l|XXXXX} \toprule
\multirowcell{2}[-20pt][l]{\textbf{Energy-intensive} \\ \textbf{industrial process}} & \textbf{Demand response potential}  & \textbf{Installed capacity Germany}  & \textbf{Max. duration of load change\tnote{a}}  & \textbf{Min. load change costs}\tnote{b} & \textbf{Load change costs} \\
     & \si{\percent} of the installed load & \si{MW_{el}} & \si{\hour} & \si{{euro} \per {MW_{el}}} & \si{{euro} \per {MW_{el}}} \\  \toprule  
     Electric arc furnace & 99\tnote{{c}} & 1097\tnote{d} & 336 & & 283\tnote{e} \\ \midrule 
     Aluminum &95\tnote{f} &543\tnote{d} &336&	&27\tnote{e}\\ \midrule 
    Paper&	95\tnote{g}	&312\tnote{d}&	12&	10\tnote{d}& 69\tnote{h}\\ \midrule 
    Cement&	90\tnote{i}	&360\tnote{d}	&336 & 80\tnote{j} &	320\tnote{e}\\  \midrule 
    Chlor-alkali process&	54\tnote{k}	&1484\tnote{l}& 72& 0\tnote{d}&	100\tnote{l}\\ \midrule 
    Air separation & 70\tnote{m}&	570\tnote{l}&	72&	0\tnote{n}&	243\tnote{m}\\ \midrule 
    Other &	10\tnote{o}&	1050\tnote{p}&	336&	10\tnote{n}&	300\tnote{n}\\ \bottomrule
\end{tabularx}

\label{tab:DR-Ind-Data}
\end{threeparttable}
\begin{tablenotes}[para,flushleft]\footnotesize \tnote{a}see assumption~6, \tnote{b}see assumption~8, \tnote{c}\cite{dena.2010}, \tnote{d}\cite{Paulus.2011}, \tnote{e}See variable load changing cost estimation calculations (equation~\ref{eq:RR}), \tnote{f}\cite{Langrock.2015}, \tnote{g}\cite{Steurer.2017}, \tnote{h}\cite{Helin.2017}, \tnote{i}\cite{VDEStudie.2012}, \tnote{j} assumption~\SI{25}{\percent} of the load change costs, \tnote{k}\cite{Klaucke.2023}, \tnote{l}\cite{Klaucke.2017}, \tnote{m}\cite{Kollmann.2015}, \tnote{n}own assumption, \tnote{o}see assumption 3, \tnote{p}see assumption~4 and~5.
\end{tablenotes}
\end{table}
\end{center}

\newpage

Integrating demand response into the production process requires initial investments and incurs both variable and fixed costs \citep{Hoffmann.2021}. To model these costs, the following simplifying assumptions are made:
\begin{enumerate}\setcounter{enumi}{6}
    \item Only the cost of activating the load change is considered as a variable cost. Variable costs resulting from the duration of the load change are neglected. 
    \item The variable cost of the load change increases with the level of activated load. For simplicity, lower costs are considered for the first  \SI{20}{\percent}~of available flexible load. This also reflects the wide range of values reported in the literature.
    \item There are no storage losses in product storage.
    \item By 2030, it is projected that existing production capacities will be underutilized as a result of the increasing relevance of the circular economy. The resulting surplus capacity can be employed for demand response overcapacity, thereby obviating the necessity for further investment costs for additional capacity expansion.
    \item Investment costs for product storage depend significantly on product properties (e.g. density, state of aggregation, hazard potential, storage conditions, etc.) and the individual site conditions of production. Therefore, the investment costs for product storage for 1,2-dichloroethane (DCE) are assumed here for all processes for the purposes of simplification. 1,2-Dichloroethane is liquid under ambient conditions and has good properties regarding storability. It is stored in a coated vessel tank \citep{Klaucke.2020}. The specific investment costs for these storage systems are estimated at~\SI{5240}{euro \per MW_{el]}} based on \cite{Klaucke.2023}. 
    \item Fixed costs of storage are neglected.
\end{enumerate}

The variable load change costs in~\si{euro \per MW} for chlor-alkali process, paper, and air separation are determined by utilizing literature data (see table~\ref{tab:DR-Ind-Data}). However, given the absence of plausible values for electric steel, aluminum electrolysis and cement, their costs were estimated by calculating the revenue requirement. In this context, this value describes the costs to be covered in case of non-production due to load reduction. It is  using the gross value added based on the work of~\cite{Hourcade.2007}. The revenue requirement ($RR_\text{x}$) for the process~$\text{x}$  is calculated using the product-specific gross value added (${GVA}_\text{x}$), the product market prices ($p_\text{x}$) and the product-specific electricity demand ($e_\text{x}$): 
\begin{equation}
    RR_\text{x} =  \frac{GVA_\text{x} \cdot p_\text{x}}{e_\text{x}}
    \label{eq:RR}
\end{equation}
Table~\ref{tab:RR-Data} provides the revenue requirement and the input data for its calculation and table~\ref{tab:DR-Ind-Data} shows the estimated load change costs.
\begin{table}[!ht]
\footnotesize
\begin{threeparttable}
\caption{Input data for calculating the costs of the load request based on the revenue requirement}
\begin{tabularx}{\linewidth}{l|XXXX} 
\toprule
\multirowcell{2}[-5pt][l]{\textbf{Energy-intensive} \\ \textbf{industrial process}}  & \textbf{Product-specific gross value\tnote{a}}   & \textbf{Product market prices}   & \textbf{Product-specific electricity demand\tnote{b}}  &\textbf{Revenue requirement} \\
     & \si{\percent} & \si{euro \per t} &  \si{{MW_{el}} \per {t}} & \si{{euro} \per {MWh_{el}}} \\  \toprule  
     Electric arc furnace & \SI{25}{\percent} & 600\tnote{c}& 0.53 & 283 \\ \midrule
     Aluminum & \SI{20}{\percent} & 2000\tnote{d} & 15 & 27\\ \midrule
     Cement& \SI{40}{\percent} & 80\tnote{e} & 0.1 & 320 \\ \bottomrule
\end{tabularx}
\begin{tablenotes}[para,flushleft]\footnotesize \tnote{a}\cite{Hourcade.2007}, \tnote{b}\cite{Paulus.2011}, \tnote{c}\cite{stahlpreise.eu.2024}, \tnote{d}\cite{finanzen.netGmbH.2024}, \tnote{e}\cite{GTAI.2024}
\end{tablenotes}
\label{tab:RR-Data}
\end{threeparttable}
\end{table}

\subsection{Demand response potential process heat}

\label{app:process_heat}
The assessment of the load-side flexibility potential for process heat for the year 2030 based on the data of \citet{Kemmler.2017}, and was conducted under the following technical assumptions:
\begin{enumerate} \setcounter{enumi}{12}
    \item The provision of process heat is facilitated by electricity, given the availability of the requisite temperature levels and process requirements. This is achieved through the utilization of a high-temperature heat pump, capable of reaching temperatures up to~\SI{160}{\degreeCelsius}, and a resistance heater, which operates within the temperature range of~\si{200{-}}\SI{500}{\degreeCelsius} \citep{Fleiter.2023}.
    \item The average coefficient of performance (COP) of the high temperature heat-pump is~\si{3{.}7} \citep{Fleiter.2023} and the efficiency factor of the resistance heater $\eta_{\text{RH}}$ is~\SI{99}{\percent} \citep{Fleiter.2023}.
     \item The aggregate installed thermal storage capacity is contingent upon the technical and economic conditions of the discrete industrial sites. The specific capacity at each site is influenced by the local heat consumers, their process requirements, and the required temperature levels. Typically, several consumers are situated together at a particular location (e.g., an industrial park), and are collectively supplied by a service provider. Each site would optimize its thermal storage size based on individual technical constraints, as well as storage and investment costs. This optimized capacity is available as a fixed value to the electricity supply system for demand response purposes. We assume an upper bound for realizing thermal storage capacity of~\SI{30}{\percent} of the process heat requirement by the year 2030 (\SI{421.2}{GWh_{th}}).
      \item The operating hours of the process heat amount to~\SI{8760}{h}.
     \item  The thermal energy storage is a packed-bed variety with a storage efficiency of~\SI{90}{\percent} \citep{Profaiser.2022b}. 
    \item The maximum thermal storage duration is~\SI{72}{h}.
   \end{enumerate}
The electrical power requirement for the heat supply of the high-temperature heat pump $P_{el,HP}^{PH}$ and the resistance heater $P_{el,RH}^{PH}$ was calculated using equation~\ref{eq:el_power_PH} and \ref{eq:el_power_RH}:
\begin{equation}
        P_{el,HP}^{PH} = \frac{Q_{20-160\,^\circ\text{C}}^{\text{PH}}}{COP}
     \label{eq:el_power_PH}
\end{equation}  
\begin{equation}
     P_{\text{el,RH}}^{\text{PH}} = \frac{Q_{160-500\,^\circ\text{C}}^{\text{PH}}}{\eta_{\text{RH}}}
     \label{eq:el_power_RH}
\end{equation}
with $Q^{\text{PH}}$ being the required thermal energy for process heating, $COP$ the average coefficient of performance (COP) of the high temperature heat-pump and $\eta_{\text{RH}}$ the efficiency factor of the resistance heater.

The maximum available storage capacity in process heat $ P_{\text{el,S}}^{\text{PH}}$ is determined by the demand for installed electrical power for process heat under assumptions 15 to 18, as well as by the maximum heat storage duration $\tau^{\text{S}}$, and  by the efficiency of the thermal storage $\eta_{\text{HS}}$:
\begin{equation}
    P_{\text{el,S}}^{\text{PH}} = 0.3 \cdot \frac{Q^{\text{PH}}}{\tau^{\text{S}} \cdot \eta_{\text{HS}}}
     \label{eq:el_power_HS}
\end{equation}

To reflect the cost of load flexibility for process heat, investment costs and flexibility costs due to heat losses over the storage period are considered. This is done under the following assumptions: 
  \begin{enumerate} \setcounter{enumi}{18}
    \item The realization of load flexibility requires an overcapacity for the process heating of~\SI{70}{\percent} of the thermal storage capacity. We consider that the specific investment costs for this overcapacity $I_{\text{EB}}^{\text{PH}}$ amount to~\SI{80}{euro \per kWh_{th}} for a resistance heater \citep{Fleiter.2023}. Any excess capacity in addition to this can be met by existing capacities, which are typically dimensioned at a higher level to cope with peak loads.
    \item The investment costs of the fixed bed thermal storage are~\SI{40}{euro\per kWh_{th}} \citep{Arnold.2019}.
    \item The storage losses as a percentage of capacity per day are \SI{3}{\percent \per d} \citep{Arnold.2019}.
\end{enumerate}
Given the efficiency of the thermal storage tank, and the efficiency factor of the  resistance heater $\eta_{\text{RH}}$, as well as assumption~19 and~20, the specific investment costs $I_{\text{DR}}^{\text{PH}}$ in~\si{euro \per kWh_{el}} for load flexibilization of the process heat are calculated as follows:
\begin{equation}
I_{\text{DR}}^{\text{PH}} = \frac{I_{\text{S}}^{\text{PH}}}{\eta_{\text{HS}} \cdot \eta_{\text{RH}}} + \frac{I_{\text{EB}}^{\text{PH}} \cdot 0.7}{\eta_{\text{RH}}}
    \label{eq:Inv_HS}
\end{equation}

\newpage

\section{Further results}

\subsection{Flexible demand capacities}

\begin{figure}[!ht]
    \centering
    \includegraphics[width=\textwidth]{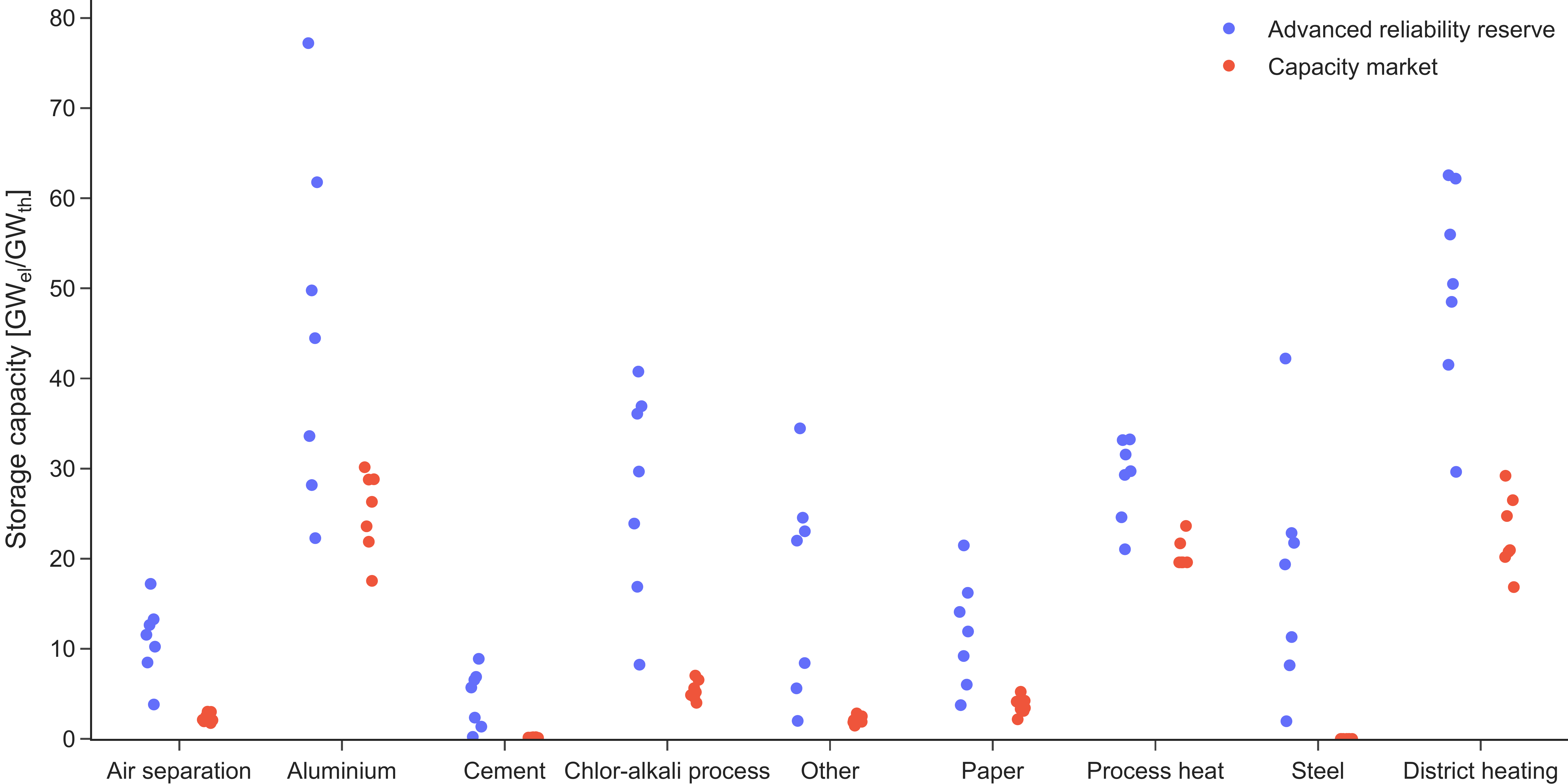}
    \begin{minipage}{\textwidth}
    \vspace{1em} \scriptsize \setstretch{1} \textit{Notes:} Storage capacity in \SI{}{GWh_{el}} for all technologies except for district heating in \SI{}{GWh_{th}}. Generation and storage capacities (of technologies in the power sector) are fixed, while the capacities of the flexible demand technologies are optimized endogenously for different weather years.
    \end{minipage}
    \caption{Optimal investment into flexibility in different weather years (2008--2014)}
    \label{fig:flex_years}
\end{figure}

As shown by many previous studies, the choice of weather year strongly affects the results of optimization models. With fixed capacities in the power sector (generators and storage), the optimal capacity size of the demand-side flexibility options varies significantly depending on the assumed weather years (Figure \ref{fig:flex_years}). The reason is that different weather years exhibit different types and lengths of periods of scarce renewable energy supply, in combination with different demand patters, which in turns leads to different optimal model outcomes. Despite the variation between weather years, the impact of the centralized capacity market and the advanced reliability reserve on optimal demand-side flexibility capacities remains the same. The latter leads to more investment.


%% file: Tables/capacities.tex
\renewcommand\tabularxcolumn[1]{m{#1}}

\begin{table}[!htbp]
\caption{Capacity and cost assumptions}
\scriptsize
\begin{tabularx}{\linewidth}{llXXXXXXX} 
\toprule 
& \textbf{Capacity}  & \textbf{Lifetime} & \textbf{Overnight costs} & \textbf{Fixed costs} & \textbf{Efficiency} & \textbf{Carbon content} & \textbf{Fuel costs} & \textbf{Var. costs storing in/out} \\ 
& GW & years & euro/kW & euro/kW & \% & euro/MWh & euro/MWh & euro/MWh \\
\toprule 
        Run-of-river hydro          & 3.93                  & 50   & 3000     & 30    & 0.9        & 0     & 0     & -        \\ \midrule
        Natural gas (CCGT)          & 0.00--$\infty$         & 25   & 800      & 20    & 0.54       & 0.201 & 26.03 & -        \\ \midrule
        Natural gas (OCGT)          & 0.00--$\infty$         & 25   & 400      & 15    & 0.4        & 0.201 & 26.03 & -        \\ \midrule
        Oil                         & 2.82                  & 25   & 400      & 7     & 0.35       & 0.266 & 41.65 & -        \\ \toprule
        Bio energy                  & 11.06                 & 30   & 1951     & 100   & 0.49       & 0     & 10    & -        \\ \midrule
        Onshore wind                & 115.00                & 25   & 1182     & 35    & 1          & 0     & 0     & -        \\ \midrule
        Offshore wind               & 30.00                 & 25   & 3935     & 100   & 1          & 0     & 0     & -        \\ \midrule
        Solar PV                    & 215.00                & 25   & 600      & 25    & 1          & 0     & 0     & -        \\ \toprule
        Lithium-ion batteries       &                       & 20   &          &       &            &       &       &          \\ 
        ... power in/out            & 0--$\infty$/0--$\infty$ &      & 50/0   & 0.1/0 & 0.97/0.97  & -     & -     & 0.3/0.3  \\ 
        ... energy {[}GWh{]}        & 0--$\infty$            &      & 300      & 0.7   & 0.999989   & -     & -     & -        \\ \midrule
        Power-to-gas-to-power       &                       & 22.5 &          &       &            &       &       &          \\ 
        ... power in/out            & 0--$\infty$/0--$\infty$ &      &  305/850 & 0/0   & 0.73/0.6   & -     & -     & 1.2/1.2   \\ 
        ... energy {[}GWh{]}        & 0--$\infty$            &      &  2       & 0     &   1         & -     & -     & -         \\ \toprule
        Open pumped hydro storage   &                       & 80   &          &       &            &       &       &           \\ 
        ... power in/out            & 1.86/2.14             &      & 550/550  & 0/0   &  0.97/0.91 & -     & -     & 0.56/0.56 \\ 
        ... energy {[}GWh{]}        & 471.23                &      & 10       & 0     &  0.999995  & -     & -     & -         \\ \midrule
        Closed pumped hydro storage &                       & 80   &          &       &            &       &       &           \\ 
        ... power in/out            & 6.56/6.41             &      & 550/550  & 0/0   &  0.97/0.91 & -     & -     & 0.56/0.56 \\ 
        ... energy {[}GWh{]}        & 391.58                &      & 10       & 0     &  0.999995  & -     & -     &   -       \\ \midrule
        Reservoirs                  &                       & 50   &          &       &            &       &       &           \\ 
        ... power out               & 0.82                  &      & 200      & 30    &  0.95      & -     & -     &  0.1      \\ 
        ... energy {[}GWh{]}        & 237.22                &      & 10       & 0     &  1         & -     & -     &  -        \\ \bottomrule
    \end{tabularx}
    \begin{minipage}{\textwidth}
    \vspace{1em} \scriptsize \setstretch{1} \textit{Notes:} In the column capacity, if two numbers are connected with a hyphen, the model can choose endogenously in that range; otherwise, the value is fixed. The assumed carbon price is~\SI{130}{euro \per ton}.
    \end{minipage}
    \label{tab:dieter}
\end{table}